# What's in a Name: History and Meanings of the Term "Big Bang"

HELGE KRAGH[*]

**Abstract.** The name "big bang" introduced by Fred Hoyle in 1949 is one of the most successful scientific neologisms ever. How did the name originate and how was it received by physicists and astronomers in the period leading up to the hot big bang consensus model in the late 1960s? How did it reflect the meanings of the big bang, a concept that predates the name by nearly two decades? The paper gives a detailed account of names and concepts associated with finite-age cosmological models from the 1920s to the 1970s. It turns out that Hoyle's celebrated name has a richer and more surprising history than commonly assumed and also that the literature on modern cosmology and its history includes many common mistakes and errors. By following the story of "big bang" a new dimension is added to the historical understanding of the emergence of modern cosmology.

> What's in a name? That which we call a rose
> By any other name would smell as sweet
> *Romeo and Juliet* (Act II, Scene 2)

## 1. Introduction: Name and histories

"Words are like harpoons," famous astrophysicist and cosmologist Fred Hoyle said in an interview of 1995. "Once they go in, they are very hard to pull out" (Horgan 1995, p. 47). He referred to the term "big bang," which he had coined nearly half a century earlier for the initial state of the universe without believing, neither then nor later, that there had ever been such an initial state. The name had indeed stuck like a harpoon, and that in spite of many people's dissatisfaction with such an undignified label for the grandest and most mysterious event in the history of the universe, the ultimate beginning of everything. Two years earlier the astronomy magazine *Sky and Telescope* had run a competition to rename the big bang theory of the universe (Ferris 1993; Beatty and Fienberg 1994). The panel of judges, consisting of astronomer Carl Sagan, television broadcaster Hugh Downs and science writer Timothy Ferris, mulled over 13,099 suggestions from 41 countries only to decide that none of them were worthy of supplanting "big bang." The name had stuck, and it still sticks.

---

[*] Centre for Science Studies, Department of Physics and Astronomy, Aarhus University, Building 1520, 8000 Aarhus, Denmark. E-mail: helge.kragh@ivs.au.dk.

Names in science are more than just names. Some are eponymous labels honouring great scientists, while others may associate a concept or an object with a particular meaning that another name would not signal. An apt name may help set the frame of mind with which scientists view an object or a concept. Names may also help making a concept more popular and widely recognized simply because they are apt, such as was the case with Murray Gell-Mann's "quark" of 1964. Contrary to names for objects such as elementary particles, chemical elements and comets, the names of theories or concepts have no official status. They are not decided by committees but just happen to be adopted by the relevant scientific community.

When a name catches on it is not only an indication of its popularity but also, in many cases, of the concept associated with the name. Because names sometimes carry epistemic connotations and mental images with them, they can be controversial (Stuewer 1986). For example, this was the case with the term "entropy" coined by Rudolf Clausius in 1865 to express the meaning of the second law of thermodynamics (namely, as he phrased it two years later, "the entropy of the universe tends to a maximum"). Clausius' entropy met with considerable resistance among leading physicists and chemists who preferred other names and concepts, such as "dissipation of heat" and "free energy" (Kragh and Weininger 1996). The eminent physical chemist Walther Nernst very much disliked the term, not only because it referred to an abstract and unnecessary concept but also because of its association with Clausius' prediction that the ever-increasing entropy would cause the world to suffer a "heat death." To Nernst, the name and concept aroused strong emotional antipathy. Over time entropy lost its controversial connotation and is today as popular as ever, used widely also in many areas outside the sciences.

Another case in point is "relativity theory," which was not coined by the originator of the theory but by Max Planck in 1906. Einstein preferred "relativity principle" and only adopted Planck's name at about 1915. When Einstein's theory became controversial in the public arena, where it was sometimes conflated with relativism, some physicists suggested renaming it "invariance theory," a name first proposed by the mathematician Felix Klein in 1910. The proposal fell on deaf ears, perhaps because of the term's similarity to the established mathematical "invariant theory." In a letter of 1921 Einstein agreed that relativity theory was in some respects a misnomer, yet it "would only give rise to confusion if after such a long time the generally accepted name were now changed" (Fölsing 1997, pp. 208-210).

It is not the aim of this essay to discuss the meanings and histories of scientific names in general, but it is worth pointing out that Hoyle's big bang is not an exceptional case. There are many other invented names that have in common with this term that they are considered catchy and, in part for this reason, have made a difference in how scientists and the public think about nature. Within modern astrophysics one may point to names such as black hole, pulsar, gamma burst, quasar, anthropic principle, dark energy and dark matter. The name and concept of "black hole" illustrates how a good name can conjure up a mental image that emphasizes the important properties of a physical concept, while other names tend to produce mental blocks that hinder recognition of these properties (Israel 1987, p. 250; Thorne 1994, pp. 254-257).

The idea of a space-time singularity created by the gravitational collapse of a massive star had been discussed within the framework of general relativity theory since the 1930s without recognizing that the word "singularity" for the critical circumference of a collapsing star (the Schwarzschild singularity) conveyed a wrong idea of what happened. Only later was it recognized that the Schwarzschild singularity was really a horizon surrounding the singular centre of the imploding star. The object was referred to as either a "collapsed star" or a "frozen star," two names that differed slightly in their meanings and both of which were unfortunate because they failed to highlight the horizon as the characteristic feature. With the gradual recognition of the importance of the horizon a new and better name was needed. John Wheeler responded to the need when he coined the name black hole in a talk of 1967 (Wheeler 1968). Apart from being catchy, Wheeler's name carried with it a different mental picture of the gravitational collapse that corresponded to the modern viewpoint of the increasingly fashionable field of black hole physics. The name was quickly adopted by physicists and astronomers as well as the general public.[1]

As illustrated by the words entropy, relativity theory and black hole, scientific names are more than just neutral labels introduced for reasons of convenience (although convenience plays a role too, of course). The momentous idea that the universe came into being in a kind of explosive act some finite time ago has for more than forty years been known as the big bang theory. The idea dates from 1931, and although Hoyle coined the name in 1949 it took more than twenty years until big bang became a household word in the scientific community. By focusing on how physicists and astronomers spoke of what became the big bang universe we can obtain complementary insight in the historical development of cosmology in the crucial period from the early 1930s to the mid-1970s. We also use the opportunity to correct some widely held misconceptions regarding both terminology and the substance of the cosmological theories in the period.

---

[1] The concept of a "white hole," a hypothetical object emerging spontaneously from a singularity – or a time-reversed version of a black hole – was introduced by Igor Novikov in 1964, but without referring to it as a white hole. The following year the concept was independently considered by Yuval Ne'eman, who called the object a "lagging core." The apt name "white hole," which may first have been used in 1971, soon became popular while "lagging core" was forgotten. Contrary to the black holes, white holes or lagging cores are not believed to exist in nature. They were sometimes called "little bangs," a term also used with somewhat different connotations (Hoyle 1965; Harrison 1968).

## 2. Early ideas of the exploding universe

Until 1931, the origin of the universe in a physical sense was not part of the new cosmology based on Einstein's field equations of 1917. There was no cosmogony in the true meaning of the term. Einstein's original model was static and hence with no beginning in time, and the same was the case with the only alternative seriously discussed in the period, a model proposed by the Dutch physicist Willem de Sitter.[2] With just two exceptions, whose contributions remained unnoticed through the 1920s, astronomers and physicists tacitly agreed that the universe was static. As late as 1929, Einstein confirmed that "the [space-time] continuum is infinite in its time-like extent," meaning that space had existed eternally (Einstein 1929, p. 108). In the few cases where a finite-age universe turned up in the literature, it was either as a philosophical speculation, a mathematical possibility or in the context of the old discussion of a cosmic heat death. In 1928, Arthur Eddington briefly considered if the state of the world could be traced back to the ultimate limit of an absolute minimum entropy. Such a state would correspond to a cosmic beginning, a concept that Eddington dismissed from both a philosophical and a scientific point of view. "As a scientist I simply do not believe that the Universe began with a bang," he said, thus inventing half of the big bang term (Eddington 1928, p. 85).

[2] The historical development of early relativistic cosmology is covered in, for example Kragh (1996) and Nussbaumer and Bieri (2009), where further information and references can be found.

The Russian physicist Alexander Friedmann is sometimes mentioned as the father or grandfather of the big bang universe. According to a recent paper, he "deserves to be called the father of Big Bang cosmology" (Belenkiy 2012). The claim rests on a paper of 1922 in which Friedmann demonstrated for the first time that Einstein's cosmological equations had dynamical solutions corresponding to, for example, an expanding universe. In this later so famous paper Friedmann even wrote about "the time since the creation of the world," namely, the time that had passed since the universe was concentrated in a point or singularity of zero volume (Friedmann 1922, p. 380). Five years later the Belgian physicist and astronomer Georges Lemaître rediscovered independently Friedmann's dynamical solutions which he compared to the new astronomical observations of galactic redshifts. He concluded that the universe was in fact expanding (which Friedmann did not), but not that it was expanding from a point-like state a definite time ago. According to the expanding 1927 Lemaître model, or what came to be known as the Lemaître-Eddington model, our universe had evolved asymptotically from a pre-existing Einstein universe and thus could not be ascribed a definite age. While the works of Friedmann and Lemaître are today recognized as foundations of modern cosmology, they were almost completely ignored until the early 1930s.

There are several standard misconceptions of the early phase in the history of the expanding universe before it developed into a big bang theory in the 1930s. First, it is often believed that the finite-age big bang universe follows as a consequence from the expanding universe, implying that the idea of a big bang is to be found, directly or indirectly, in Lemaître's 1927 paper (e.g., Singh 2004, p. 158; Marx and Bornmann 2010, p. 450; Raychaudhury 2004).[3] This is just wrong. Not only is there no necessary path from the expanding to the exploding universe, such as is illustrated by the Lemaître-Eddington model; the idea of a big bang was also absent from Lemaître's paper of 1927 and only appeared four years later (Kragh and Lambert 2007). A second misunderstanding is to read the big bang into Friedmann's paper of 1922. Although this paper included big bang solutions, it was in a mathematical sense only ($R = 0$ for $t = 0$) and without the Russian physicist paying more attention to these solutions than to other solutions. In fact, his paper was basically mathematical, with no references to either astronomical measurements (such as the galactic redshifts) or to the physical properties of the universe.

---

[3] This standard error also appears in the *Oxford English Dictionary* (online edition of 2012, entry: big bang) which dates Lemaître's primeval atom hypothesis to 1927.

Curiously, none other than George Gamow, perhaps more than anyone the true father of the big bang, credited Friedmann with having introduced the hot big bang model. Gamow, who in his youth had studied under Friedmann, wrote as follows: "According to Friedmann's original theory of the expanding universe, it must have started with a 'singular state' at which the density and temperature of matter were practically infinite" (Gamow 1970, p. 141; Chernin 1995). This is however Gamow reading his own theory into the one of his former teacher. Friedmann's theory of 1922 had nothing to say about the density and temperature of the early universe or of any of its physical properties.[4]

In a brief letter to *Nature* of 9 May 1931 – a date that can reasonably be seen as the birth of the big bang universe – Lemaître suggested that the universe had come into being in an explosive act governed by the laws of quantum mechanics. "We could conceive the beginning of the universe in the form of a unique atom," he wrote, likening the original state to a huge atomic nucleus that would decay or explode by a "super-radioactive process" and thereby initiate the cosmic expansion (Lemaître 1931, reprinted in Lemaître 1950, pp. 17-19). The note in *Nature* was a qualitative and imaginative hypothesis, a visionary piece of cosmo-poetry in which Lemaître expressed in words what could not be expressed in words. Half a year later he turned the poetry into a proper scientific theory framed in the language of relativistic cosmology, and at the same time he discussed it at a meeting in London celebrating the centenary of the British Association for the Advancement of Science.

---

[4] Trying to understand the reason for Gamow's imaginative account of Friedmann's theory, Chernin (1995, p. 451) suggests that the idea of an almost infinite density and temperature near the singular state was considered "natural or even trivial by Friedmann himself and the people around him." However, this is clearly an anachronistic interpretation that lacks documentation.

Lemaître's model was of the big bang type earlier considered by Friedmann, but the Belgian cosmologist was careful to adopt a physical rather than mathematical point of view. According to his thinking there was no initial zero-volume singularity corresponding to an infinite density, but instead everything began in a material proto-universe of a density of the order of an atomic nucleus and a size comparable to the solar system. This primordial atom or nucleus was changeless and completely undifferentiated, devoid of physical qualities. Although Lemaître stressed that it was physically real, he also said that it was inaccessible to science, for "in absolute simplicity, no physical questions can be raised" (Lemaître 1958, p. 6). Structure and change only started with the radioactive explosion with which the universe came into being and time began. As to the origin of the primeval atom itself he was silent, but there is little doubt that Lemaître, who was a Catholic priest, believed it was created by God.

To describe in words the original state of the universe, Lemaître had to take recourse to a metaphorical terminology. In his note to *Nature* he spoke of an "original quantum" and a "unique atom," while in later publications his preferred name was the "primeval atom" (*l'atome primitif* in French). His only book on the subject, a collection of articles published in 1946, was entitled *L'Hypothèse de l'Atome Primitif* and translated into English in 1950 as *The Primeval Atom*. On some later occasions he used for the primeval atom the image of a base nucleus without surrounding electrons, "a kind of isotope of the neutron" (Lemaître 1946, p. 155). At the London 1931 meeting he also used another metaphor, now referring to the "fireworks theory" of cosmic evolution. He realized of course that the apt metaphor was flawed, as fireworks explode into the surrounding space whereas there was no space into which the primeval atom-universe could expand.

In his publications on cosmology from 1931 to the early 1960s Lemaître usually referred to his theory by these two names, primeval atom and fireworks. He never used the term big bang, neither for his own theory nor for other theories of the early universe. In the modern literature one can find several references to Lemaître's "cosmic egg," sometimes with the information, given in quotation marks but always without a source, that Lemaître described his primeval atom as "the Cosmic Egg exploding at the moment of the creation" (Sidhart and Joseph 2010, p. 642; Trimble 2000, p. 4). The first reference to Lemaître's cosmic egg may have been in John D. Bernal's *Science in History* of 1954, which included a brief section on the controversy between evolution and steady state theories in cosmology. According to Bernal (p. 541): "Lemaître in 1927 [*sic*] made the drastic assumption that all the matter in the universe was packed into one atom, a kind of cosmic egg, which burst in the first and greatest atomic explosion, not four thousand but four thousand million years ago." Two years later, John Pfeiffer, in a popular book on the new science of radio astronomy, referred repeatedly to the cosmic egg theory rather than using the equivalent term big bang theory (Pfeiffer 1956, pp. 225-229). Again, in 1962 astronomers Otto Struve and Velta Zebergs mentioned "something G. Lemaître has described as the 'primeval egg' of the universe" and which "is related to what astronomers have called the 'big-bang' theory of cosmology" (Struve and Zebergs 1962, p. 472). Whatever the origin of the egg metaphor, it is doubtful if Lemaître actually used it, although he may have done so informally in conversations.

Lemaître's theory of the beginning of the universe was not warmly welcomed by the majority of astronomers and physicists, who tended to ignore it or dismiss it as a speculative scenario unsupported by facts. It was generally considered "a clever *jeu d'esprit*," as one critic called it (Barnes 1933, p. 408). On the other hand, the primeval atom hypothesis was received with much interest in newspapers and popular science magazines. In 1932 *Popular Science* included a paper dramatically entitled "Blast of Giant Atom Created Our Universe" in which the Harvard astronomer Donald Menzel described Lemaître's theory in some detail. The essence of the theory – that "the whole universe was born in the flash of a cosmic sky rocket" – was described by the use of images relating to explosions and fireworks (Menzel 1932). Among the very few scientists in the 1930s who expressed some sympathy for finite-age models of Lemaître's type, the German quantum physicist Pascual Jordan may have been the only one who explicitly referred to Lemaître. Also Jordan used the explosion metaphor to describe how the very early universe had arisen from an "original explosion" or in German an *Urexplosion* (Jordan 1936, p. 152; Jordan 1937). The commonly used German term for the big bang is *Urknall*, meaning the original or primordial bang. A kind of *Urknall* was at the same time considered by the German physicist and meteorologist Hans Ertel, who apparently came to the idea independently of Lemaître (Schröder and Treder 1996).

## 4. A big bang universe without a name

The Russian-American nuclear physicist George Gamow is as strong a candidate for paternity of the big bang theory as Lemaître, and that even though he came to the idea many years later and by following a very different approach (Kragh 1996, pp. 101-141; Harper 2001). This approach focused on how the chemical elements were formed, which according to Gamow's hypothesis had occurred in nuclear processes during of brief period of time in the hot and dense early universe. As early as 1940, in his popular book *The Birth and Death of the Sun*, he suggested that the naturally occurring radioactive elements had been formed shortly after "the creation of the universe from the primordial superdense gas" (Gamow 1940, p. 201). Six years later he combined the idea with the Friedmann equations, and over the next few years he developed it in collaboration with his assistants Ralph Alpher and Robert Herman. What I shall for simplicity call the Gamow theory or model of the early universe presupposed an initial state of highly compressed and very hot nuclear matter, at first taken to consist of neutrons only but subsequently to be revised to a mixture of neutrons and protons. Another important insight was that in the earliest phase of the expanding universe, when the temperature was exceedingly high, it was dominated by electromagnetic radiation rather than matter.[5]

      Neither Gamow nor his two associates attempted to explain how the original nuclear substance had come into existence, they just took it for given. Their theory was an evolution theory, not a creation theory. They sometimes referred to the primordial soup of photons and nuclear particles as "ylem," an ancient word for the original matter of the world that was first suggested by Alpher in a paper of 1948. Gamow quickly appropriated the term and used is frequently in his writings, but it was rarely used by other physicists. What matters is that by 1950 Gamow, Alpher and Herman had developed a sophisticated theory of the early universe based on nuclear-physical calculations and the relativistic standard theory of the expanding universe. Although they did not succeed in explaining the formation of the heavier elements on this basis, they did account for the abundance of helium in the universe.

      Although Gamow's theory was well known in the early 1950s, it was accepted only by a small minority of physicists (and by no astronomers). Rather than associated with a particular word or phrase, people used a variety of names for it,

---

[5] As Alpher and Herman realized in 1948, there would still be a relic of the original blackbody radiation, now at very low temperature ($\approx 5$ K) and predominantly in the microwave region. They predicted the cosmic microwave background that was independently discovered in 1964 and then changed the course of cosmology. However, at the time of their prediction it was ignored (Alpher and Herman 1990).

none of which can be characterized as catchy. Hoyle's "big bang" was occasionally used in the period 1950-1965, but the term did not refer specifically to Gamow's theory but most often to finite-age models in general. Another name was the "αβγ theory," a reference to an important 1948 paper nominally written by Alpher, Hans Bethe and Gamow. (Meant as a joke, Gamow added Bethe as an author without his knowledge.) Gamow himself characterized his theory by inapt names such as "the non-equilibrium theory of the building up of nuclei," "the relativistic evolution theory," and "the hypothesis of 'beginning'." The Russian physicist Yuri Smirnov referred to the theory as the "Gamow model" or the "theory of prestellar evolution" (Smirnov 1965, translation of Russian article of 1964).

In a popular book of 1952 with the then provocative title *The Creation of the Universe* Gamow gave a full account of his big bang theory without using the term. On the other hand, he coined the name "big squeeze" for the collapsing universe that might be imagined to have preceded the present expansion. Interestingly, Gamow sometimes referred to the big squeeze in ways that were almost indistinguishable from the big bang. He thus spoke of "the Big Squeeze which took place in the early history of our universe" (Gamow 1952, p. 36) and took the date of the big squeeze to be equal to the age of the universe. [6]

Given that Gamow's universe was in fact of the big bang type, it is perhaps understandable that some authors have mistakenly believed that he was the originator of the name. The Swedish physicist Hannes Alfvén, a Nobel Prize winner of 1970 and a sharp critic of what he considered the myth of the big bang, was perhaps the first to convey the mistake, to which he added another one. In a book of 1966, he wrote: "Gamow introduced a number of suggestive terms, among them 'ylem' for the dense primeval clump of matter and 'big bang' for the explosion itself" (Alfvén 1966, p. 15). Again, in a widely read book by Timothy Ferris: "Gamow … authored the modern approach to Lemaître's expanding-universe cosmology, for which he coined the term 'Big Bang theory'" (Ferris 1977, p. 77).[7] The truth is that Gamow resented the name big bang, which he almost never used. As far as I know, he only mentioned it once, in a popular review article in *Scientific American* (Gamow 1961). Neither did Alpher and Herman refer to the big bang during their scientific careers. When they eventually did, in a historical review paper of 1969, it was with "big bang" in inverted commas (Alpher and Herman 1969).

---

[6] In later literature the names "big squeeze" or "big crunch" were mostly used for the ultimate end of the universe in the far future, should the gravitational contraction overpower the force of expansion. Such a scenario presupposes that our universe is closed, which Gamow denied. "Big crunch" may today be more common that "big squeeze." Whereas the former has an entry in the *Oxford English Dictionary*, the latter has not.

In an interview of 1968 Charles Weiner mentioned to Gamow that in the popular mind his name and the big bang were almost synonymous. Gamow agreed, but was not pleased. "I don't like the word 'big bang'," he said. "I don't call it 'big bang,' because it is a kind of cliché. This was invented, I think, by steady-state cosmologists – 'big bang' and also the 'fire ball' they call it, which has nothing to do with it – it's not fire ball at all. Nothing to do with the fire ball of atomic bomb" (AIP 1968). Gamow was of course aware of the name Hoyle had proposed for the kind of theory that contrasted so strongly with Hoyle's own steady state theory. In the late 1950s radio-astronomical measurements made by Martin Ryle and his group in Cambridge seemed to contradict the steady state theory, which Ryle strongly disliked. In part as a result of their disagreements about the interpretation of the radio data, the relationship between the two distinguished astronomers evolved into a major feud between them. The controversy caused Barbara Gamow, the wife of

---

[7] The same mistake appears in Peebles (1984, p. 26). See also Mitton (2005, p. 129), who wrongly says that Gamow promoted the name big bang.

George Gamow, to describe in the form of a poem an imagined discussion between Ryle and Hoyle.[8] In two of the verses Hoyle speaks to Ryle (AIP 1968).

> Said Hoyle, "You quote
> Lemaître, I note,
> And Gamow. Well, forget them!
> That errant gang
> And their Big Bang –
> Why aid them and abet them?

> You see, my friend,
> It has no end
> And there was no beginning
> And Bondi, Gold,
> And I will hold
> Until our hair is thinning!"

## 4. Hoyle's big bang – not a big deal

The steady state theory of the universe was proposed in 1948 in two different versions, the one by Hoyle and the other by his friends Hermann Bondi and Tommy Gold (Kragh 1996). According to this theory, although the universe was expanding, on a large scale it had always looked the same and would remain to do so. The universe satisfied the "perfect cosmological principle," meaning that there was no privileged place and no privileged time either. To make the constant average density of matter agree with the expansion, Hoyle and his two colleagues introduced the radical hypothesis of a continual and spontaneous creation of matter throughout space, although at a rate so small that it would not be directly detectable (namely, about $10^{-43}$ g/cm$^3$ per second).

---

[8]  Another piece of poetry, relating to the Hoyle-Ryle clash, was due to US columnist Art Buchwald: "Said Ryle to Hoyle / 'Please do not boil / The World began with a bang' / Said Hoyle to Ryle / 'Well boil my bile / Your theory doesn't hang'." *New York Herald Tribune*, February 1961, as quoted in Gregory (2005, p. 131).

Having no beginning and no end in time, obviously the Hoyle-Bondi-Gold universe was in strong contrast to the finite-age evolution theories of, for example, Lemaître and Gamow. The postulated creation of cosmic matter, assumed to be in the form of hydrogen atoms, was controversial and often seen as the characteristic feature of the theory. For this reason the steady state theory was sometimes described as the new "creation theory," a term used by Hoyle, among others (e.g. Hoyle 1949; McVittie 1951). In the later literature creation cosmology would typically refer to the big bang theory.

On 28 March 1949 Hoyle gave a twenty-minute's talk on the new cosmological theory to BBC's Third Programme (Mitton 2005, pp. 125-135; Gregory 2005, pp. 46-53). Less than two weeks later the text was reproduced in full in *The Listener*, the widely sold BBC magazine.[9] Having described the essence of the steady state theory, he contrasted it to theories based on "the hypothesis that all the matter of the universe was created in one big bang at a particular time in the remote past" (Hoyle 1949, p. 568). This is the origin of the cosmological term "big bang," which Hoyle mentioned three times. At the end of his talk, Hoyle made it clear that he found this kind of theory unacceptable on both scientific and philosophical grounds, in particular because the big bang creation process was "irrational" and outside science. "I cannot see any good reason for preferring the big bang idea," he concluded.

Later the same year Hoyle was again approached by the BBC, this time to give a series of five broadcasts on *The Nature of the Universe*. Hoyle accepted and his radio talks were on the air in January and February 1950. Half a year later the Third

---

[9]  Hoyle's original typescript is in the Cambridge University Library and can be found online on http://www.joh.cam.ac.uk/library/special_collections/hoyle/exhibition/radio/

Programme broadcasts were repeated with only superficial changes on BBC's more popular Home Service programme. The broadcasts were a smashing success, not only in England but also in Canada, Australia and elsewhere in the former British Empire. They were printed verbatim in *The Listener* and within a few months they appeared in book form, published in England by Blackwell and in the United States by Harper & Brothers. The books followed the original scripts closely, with the American edition differing from the British only by minor changes and by incorporating the notes in the text.

Hoyle referred twice to the big bang in the British edition of *The Nature of the Universe* (on p. 102 and p. 105), and in ways which were not clearly pejorative or derisive. He described the big bang creation of the universe as "much queerer than continuous creation," but not as "irrational" as he had done in his 1949 broadcast and as he also did in the American edition, where the term big bang appeared four times (Hoyle 1950b, p. 105; Hoyle 1950a, p. 124). When the book was republished in 1960, Hoyle included the somewhat stronger wording of the American edition. There were many critical responses to the BBC broadcasts and the book that followed, but very few of them paid attention to the name for the exploding universe that Hoyle had so casually invented. For example, the name was not mentioned in Williamson (1951) and O'Connell (1952), two highly critical reviews written by astronomers.[10]

There are in the literature some misconceptions about Hoyle's BBC addresses, including Gamow's possible role in them. According to one account, Hoyle "jokingly referred to Gamow's theory of the creation of the universe as the 'Big Bang'" (Raychaudhuri 2004, p. 37). In fact, Hoyle did not mention Gamow or his theory at all, and when he referred to Lemaître, which he did once, it was not to his primeval-atom hypothesis. In a symposium on the history of cosmology held in Bologna in 1988, Alpher and Herman, who were Gamow's closest associates in the years around 1950, reported: "According to Gamow, Hoyle first used this phrase [big bang] in a pejorative sense during a BBC radio debate with Gamow" (Alpher and Herman 1990, p. 135). And a few years later, now stating it as a fact: "Toward the end of 1949 Gamow engaged in a transatlantic debate with Hoyle on BBC. It was during this debate that Hoyle first used the designation 'Big Bang,' and in a pejorative sense" (Alpher and Herman 1997, p. 63). There never was such a radio debate between the two cosmologists! Perhaps Alpher and Herman misunderstood something Gamow had told him, or perhaps Gamow just told the story as a joke (which would have been in the spirit of the pun-loving Gamow).

---

[10]  I know of only two exceptions. A book review in *Science* mentioned Hoyle's dislike of the "big bang idea" (Mather 1951), and a similar reference appeared in an editorial comment in *Popular Science* (vol. 159, July 1951, p. 78) which promoted Hoyle's book by printing a chapter from it. Whereas the name "big bang" aroused little interest among either scientists or the general public, Hoyle's critical comments on religion in the last part of his book gave rise to a heated debate (Gregory 2005; McConnell 2006). In this debate, which was largely restricted to Great Britain, Hoyle's name for the exploding universe played no role.

It is possible that Gamow felt Hoyle's "big bang" to be a pejorative phrase, but there is no documentation that either Gamow, Lemaître or other protagonists of explosion cosmologies at the time felt offended by the term – or that they paid attention to it at all. In any case, with the later success of the big bang theory it became common to see Hoyle's neologism as an attempt to make the idea of an explosion universe, and Gamow's version of it in particular, sound ridiculous. This is not how Hoyle saw it. At the time he seems to have considered it just an apt but innocent phrase for a theory he was opposed to, and he later insisted that he had not thought of it in a derogatory sense. In an interview of 1989, Alan Lightman asked him if he was really the source of the name. Hoyle answered (Lightman and Brawer 1990, p. 60):

> Well, I don't know whether that's correct, but nobody has challenged it, and I would have thought that if it were incorrect somebody would have said so. I was constantly striving over the radio – where I had no visual aids, nothing except the spoken word – for visual images. And that seemed to be one way of distinguishing between the steady-state and the explosive big bang. And so that was the language I used.[11]

As a broadcaster Hoyle needed word pictures to get over technical and conceptual points, and "big bang" was one of them. When he had to explain the expansion of the universe, he made extensive use of the inflating-balloon image that had first been introduced by Eddington in 1931 to illustrate a positively curved space with increasing radius.[12] Likewise, to illustrate the slow rate of matter creation in the steady state theory, Hoyle appealed to pictures familiar to all Britons. In his first BBC broadcast he explained that it would take about one billion years until a new atom was created in "a volume equal to a pint of milk bottle" (Hoyle 1949, p. 568). And the next year he said about the creation rate that it was "no more than the creation of one atom in the course of about a year in a volume equal to St. Paul's Cathedral" (Hoyle 1950a, p. 106).[13]

The standard view, to be found in numerous books and articles, is the one reported in a popular book by Nobel laureate astrophysicist George Smoot and his coauthor, science writer Keay Davidson. "Hoyle had meant the term [big bang] to be derogatory," they say, "but it was so compelling, so stirring of the imagination, that it stuck, but without the negative overtones" (Smoot and Davidson 1994, p. 68). As we shall see, it took twenty years until the term was seen as compelling and stirring of the imagination.

---

[11] Interview with Fred Hoyle by Alan Lightman, 15 August 1989, Niels Bohr Library & Archives, American Institute of Physics. The interview can be found online on http://www.aip.org/history/ohilist/34366.html. See also Horgan (1995).

[12] Eddington (1931, p. 415) invited his readers "to imagine the nebulæ to be embedded in the surface of a rubber balloon which is being inflated."

[13] Realizing that American readers might not be familiar with St. Paul's cathedral, in the American edition Hoyle referred instead to "a moderate-sized skyscraper" (Hoyle 1950b, p. 125).

## 5. Non-cosmological contexts

Given that a "bang" typically refers to an ordinary explosion, one should not be too surprised to read of big bangs also in non-cosmological contexts. In fact, although Hoyle coined the term in the cosmological meaning it is used today, he was not the first scientist to speak of a big bang. In the 1920s meteorologists and atmospheric physicists studied large explosions to learn more about the propagation of sound waves and the constitution of the atmosphere. "Methods of producing the 'Big Bang' … may not be the most effective for the purpose," we read in a note of 1924, and two years later the leading British geophysicist Francis Whipple similarly stressed the advantage of frequent smaller explosions as compared with "the occasional 'big bang'" (Ryle 1924; Whipple 1926, p. 313). The first scientific paper with "big bang" in its title was a meteorological analysis of the winds caused by a 5,000-ton TNT explosion that occurred in the spring of 1947 on the island of Helgoland (Cox et al. 1949). The paper was received by the *Journal of Meteorology* two months before Hoyle coined his memorable phrase.

Eight years later, at the height of the controversy over the steady state universe, we have again "big bang" in a headline, but also in this case without any connection to cosmology. What was called the big bang in the weekly magazine *The Economist* was a reference to the atomic bomb, and more specifically to the British plan of testing a hydrogen bomb (Economist 1957). The Cold War nuclear weapons race was also the context in which the eminent British-American theoretical physicist Freeman Dyson made a non-cosmological reference to the big bang. Referring to the difference between the fission bombs and the thermonuclear fusion bombs, he noted that "it is relatively much cheaper to make a big bang than a small one" (Dyson 1959,

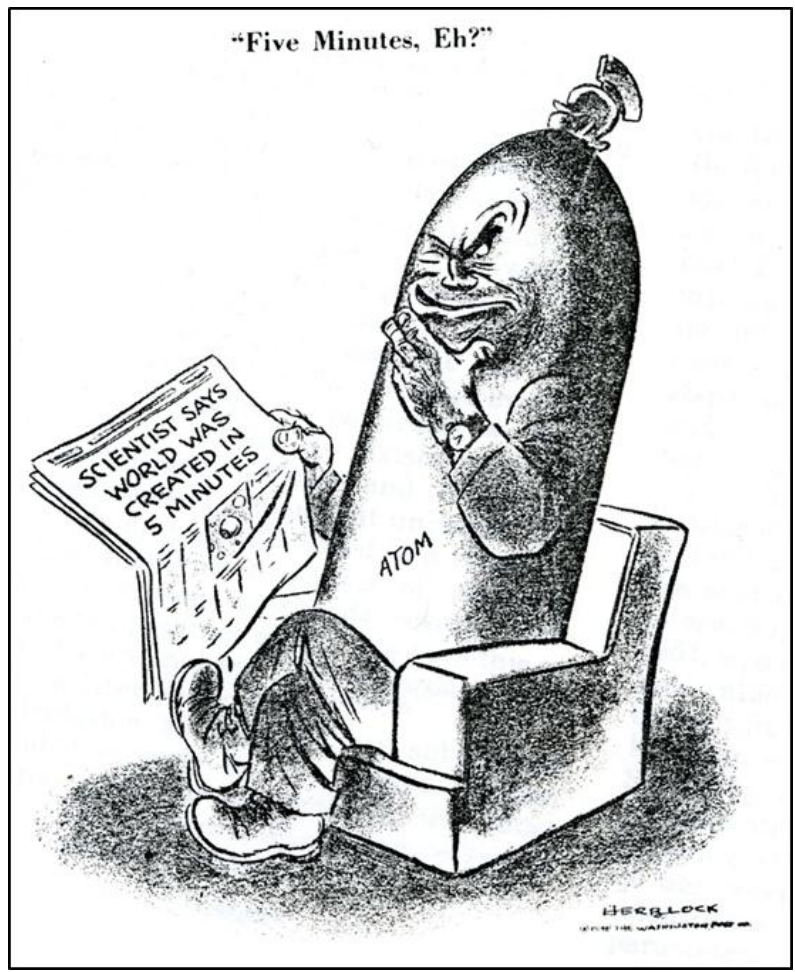

Figure 1. Herblock's cartoon alluding to the big bang origin of the world and the danger of nuclear warfare destroying it. Reproduced from Alpher and Herman 1988, p. 28.

p. 458). The use of big bang as a referent to massively destructive weapons, such as nuclear bombs, was fairly common in the 1950s. "After the technological revolution in the twentieth century, God is on the side of the big bang," one paper said, and it was not referring to God as the creator of the universe (Baldwin 1956, p. 153). The term was often used in relation to nuclear warfare, both before and after the acceptance of big bang cosmology (Schneider 1975). But it could also refer to explosions of a much smaller scale, for example to powerful cartridges for rifles (McCafferty 1963). According to David Kaiser (2007), when the phrase "big bang" appeared in American newspaper during the 1940s and 1950s, in the large majority of cases it referred to nuclear weapons and not to the origin of the universe.

We find the same connotation in John Osborne's play *Look Back in Anger*, which was first performed in 1956 and three years later turned into a movie with Richard Burton in the role as Jimmy Porter, a young disaffected man of working class origin. Jimmy says: "If the big bang does come, and we all get killed off, it won't be in aid of the old-fashioned, grand design. It will just be … as pointless and inglorious as stepping in front of a bus" (Act III, Scene 1). Although Osborne did not have cosmology in mind, and most likely was unaware of Hoyle's earlier coining of "big bang," during the Cold War it was not unnatural to associate the new atomic bombs with the hypothesis of an exploding universe. As early as 16 April 1948 *The Washington Post* included an article reporting on Alpher's doctoral dissertation on element formation in the very early universe. Headlined "World Began in 5 Minutes, New Theory," the article was accompanied by a cartoon in which Herbert Lawrence Block ("Herblock") presented an evil-looking atomic bomb contemplating if five minutes would be enough to destroy the world (Figure 1).

Also *Science News Letter* of 24 April 1948 included a reference to the atomic bomb in its account of Alpher's new creation theory of the universe, which it summarized as follows (SNL 1948):

> At the very beginning of everything, the universe had infinite density concentrated in a single zero point. Then just 300 seconds – five minutes – after the start of everything, there was a rapid expansion and cooling of the primordial matter. The neutrons – those are the particles that trigger the atomic bomb – started decaying into protons and building up the heavier chemical elements. … This act of creation of the chemical elements took the surprisingly short time of an hour. (The Bible story said something about six days for the act of creation.)

It can be argued that big bang cosmology resonated with the atomic age, whereas steady state cosmology had no such association, and that this was a contributing reason for the popularity of big bang theories in the United States in particular. There probably was some kind of link in the public mind, as illustrated by a 1950 article in *The Observer* according to which "The present popular interest in cosmology seems to have a connection with the atomic bomb" (Gregory 2005, p. 73). On the other hand, the connection is invisible in the scientific literature where theories of the big bang type were far from popular. In fact, in the period from 1954 to 1964 they were largely absent.

## 6. A catchy name that didn't catch on

According to Marx and Bornmann (2008, p. 454), "the catchy phrase [big bang] … caught on, in both camps." In reality, though, it was pretty much ignored and only appeared insignificantly in the scientific literature until the 1970s. The cosmological controversy in the 1950s is often misunderstood to be between the steady state theory and the big bang theory, or even personalized to be a fight between Hoyle and Gamow. We have an early example of this dramatized version in a comment in *Popular Science* of March 1962, portraying the controversy as a fight between the two scientists (Mann 1962, p. 29):

> In one corner you have burly, pun-making Russian-American physicist George Gamow. He says the universe did have a beginning and that beginning was a very big bang. About 10 million years ago all the atoms that now make up the uncountable stars were concentrated into one point, a superheavy glob. This glob exploded suddenly, throwing matter throughout the vastness of space. … In the other corner you have piano-playing, novel-writing, baggy-tweeded English astronomer Fred Hoyle. His side says that there was no instant creation. The universe is in a steady state. Hydrogen

atoms – the simplest form of ordinary matter – are always being generated throughout space. They always were and always will be.

However, Gamow's theory of the early universe (or rather the Gamow-Alpher-Herman theory) played very little role in the predominantly British debate. The only time that Hoyle and Gamow engaged in a kind of direct confrontation was in 1956, when both scientists argued their case in companion articles in a special issue of *Scientific American* (Gamow 1956; Hoyle 1956). What the steady state theory challenged was not specifically physical big bang conceptions of either Gamow's or Lemaître's type, but more generally the class of evolution theories based on the Friedmann equations, and especially those with a singular beginning (Deunk 1962). Hoyle, Bondi and Gold often mentioned the Einstein-de Sitter model of 1932 as the exemplar of the point-source models they were opposed to (e.g., Hoyle 1961). The Einstein-de Sitter model was of the big bang type in so far that it was expanding from a definite origin at $R = 0$ for $t = 0$, but there was no physical explosion or big bang built into the model.

The number of references to the cosmological big bang before the discovery in 1965 of the cosmic microwave background seems to have been small, restricted to just a few dozens of which most were in the American popular literature.[14] Nonetheless, there were several papers, notes and books in which the term appeared (Deunk 1962). I have counted 34 sources in English language that mention the name before 1965, and it is likely that there are several more. Of these sources, 23 are of a popular or general nature, while 7 are scientific contributions and 4 are philosophical studies. The national distribution is USA (24), Great Britain (9), Germany (1) and

---

[14]  The *Web of Knowledge* lists no papers in the period 1950-1965 with "big bang" in their titles (http://wokinfo.com). Nor does the very extensive SAO/NASA Astrophysics Data System reveal astronomical literature with abstracts including the term (hhtp://adsabs.harvard.edu/). The *Scopus* database starts only in 1960 and lists no big bang titles before 1966 (http://www.scopus.com).

Australia (1). Hoyle often compared the steady state theory with cosmologies of the big bang type, but without using the term he had invented. He preferred to speak of, for example, "Cosmologies … which predict that a definite origin of the whole universe occurred a precise finite time ago," as he phrased it in lecture delivered before the London Physical Society in 1960 (Hoyle 1961, p. 1).

The first time the term appeared in a research publication may have been in a paper of early 1957 by William Fowler, a nuclear physicist at Caltech's Kellogg Laboratory and a future Nobel laureate. Fowler was well acquainted with Hoyle, who had stayed at Caltech in 1953 and again in 1956, working on the theory of element formation in the stars. Together with Hoyle and Margaret and Geoffrey Burbidge, Fowler developed in 1956-1957 a comprehensive and soon famous theory of stellar nucleosynthesis known colloquially as the $B^2HF$ theory. In his paper in *Scientific Monthly* Fowler also considered the alternative favoured by Gamow, namely that the elements had been formed in the very early hot universe. In this connection he mentioned the postulate of "a primordial 'big bang' in which all the matter of our universe was ejected with high velocity from a common region" (Fowler 1957, p. 89). He most likely had come to the name by his close contact with Hoyle. In an earlier and more popular version, Fowler said that Gamow's theory assumed "that the cosmos started from a core which exploded in a primordial 'big bang' some five billion years ago" (Fowler 1956, p. 21).

Another of Hoyle's collaborators was the English astronomer Raymond Lyttleton, who much preferred the steady state theory over the evolutionary theories of the kind proposed by Lemaître and Gamow. In a popular book of 1956 based on a series of broadcasts in BBC's television program, he referred to "the 'big bang' hypothesis," which he compared unfavourably to the steady state theory of Hoyle, Bondi and Gold (Lyttleton 1956, p. 197). Three years later, in a review of Bernard Lovell's *The Individual and the Universe*, a published version of the BBC Reith Lectures of 1958, Lyttleton expressed his dissatisfaction with Lovell's sympathetic account of "the 'big-bang' exploding super-atom hypothesis" which to his mind was nothing but metaphysics (Lyttleton 1959). Lovell did not himself use the term big bang in his book. Among the British supporters of the steady state model was also William McCrea, an authority in cosmology and relativity physics. In a talk given to a 1962 summer school on radio astronomy he briefly referred to the big bang as a name for the class of evolutionary theories that assumed a singular initial state (McCrea 1963, p. 216).

At a 1961 conference in Santa Barbara, California, the German astronomer and leading cosmologist Otto Heckmann advocated a model in which the whole universe was slowly rotating. The advantage of this model, he thought, was that the initial state of infinite density, the singular big bang, could then be avoided. Referring four times to the big bang, Heckmann argued that it would be "incorrect to conclude from the Hubble law that there must have been a 'big bang' in the strict sense" (Heckmann 1961, p. 603). He apparently associated the term with the singularity of infinite density at $t = 0$, not with the explosion of Lemaître's extended primeval atom. Lemaître, who attended the conference, agreed that the initial singularity was an idealization. Without referring to the name "big bang," he maintained his belief in a primeval atom as the original state of the world. The following year 28-year-old Steven Weinberg published a paper on the cosmological role of neutrinos in which he referred to the "big bang theory," undoubtedly the first time the name appeared in *Physical Review* (Weinberg 1962). As he mentioned in a footnote, his investigation of neutrinos in big bang cosmologies was indebted to a personal communication from Hoyle.

The question of the origin of the universe was discussed by George McVittie, an eminent British-American astronomer and cosmologist of an empiricist inclination. McVittie was not only a staunch opponent of the steady state theory, he also disliked theories that included a sudden beginning of the universe. He described himself as belonging to the "observational school," namely, the approach to cosmology that "does not aspire to finality but only to discovering what model best fits the data" (McVittie 1951, p. 71; Sanchez-Ron 2005). Following this approach he saw no reason to accept a universe created in a big bang. In a critical and careful book review essay of 1961 McVittie objected to the idea of a big bang as apparently justified by the Friedmann equations.[15] Some of the solutions to these equations led in a formal sense to the consequence that at $t = 0$ all matter in the universe was concentrated at a single mathematical point, but was this proof of a big bang? Not according to McVittie (1961, p. 1232):

> It is said that certain models of the universe deduced from general relativity involve an initial "nuclear explosion" or a "big bang" which initiates the start of the expansion. … General relativity predicts no nuclear explosion, big bang, or instantaneous creation, for that matter, as the cause of the start of the expansion at that moment. Such notions have been woven round the predictions of general relativity by imaginative writers.

McVittie did not identify the imaginative writers, but he most likely had Gamow and Lemaître in mind. In a later paper, written at a time when the hot big bang had become the standard model, he amplified his critique of the physical big bang and its wide use in "semi-popular expositions of cosmology" (McVittie 1974, p. 260). He thought the term was "loaded with inappropriate connotations" such as an exploding bomb, and for this and other reasons "it is unfortunate that the term 'big bang', so casually introduced by Hoyle, has acquired the vogue which it has achieved."

Not unlike McVittie, the British physicist and cosmologist William Bonnor was as critical to the steady state theory as he was to big bang theories, suspecting that both classes of theories violated established standards of science. In a popular book completed in the summer of 1963, he included a section headlined "The Big Bang" in which he discussed "The start of the expansion [that] is colloquially referred to as the 'big bang'." The subsequent section was entitled "Was there Really a Big Bang?" a question he answered with a resounding no. Bonnor rejected the idea of a big bang, in part for philosophical reasons and in part because "neither the fusion-bomb theory of Gamow nor the primeval fission-bomb of Lemaître is successful in accounting for the origin and abundance of the heavy elements" (Bonnor 1964, p. 112 and p. 115). Rather than choosing the steady state theory as an alternative, he argued in favour of an ever-oscillating universe as described by the equations of general relativity. The American science writer Martin Gardner may have been the first to refer to the big bang in a book chapter headline ("Big Bang or Steady State?"). Contrary to most other writers at the time, he referred to the term without putting it in quotation marks (Gardner 1962, pp. 167-179).

---

[15]  McVittie reviewed four popular cosmology books, among them the second edition of Hoyle's *The Nature of the Universe*, which he compared in detail to the first edition.

Astronomers and physicists were not the only ones to make sporadic use of the term big bang before 1965. It turned up in the philosophical literature as early as 1953 (Hospers 1953, p. 350). Norwood Russell Hanson, a distinguished philosopher of science at Yale University, apparently liked the term which he used repeatedly in an analysis of the concept of creation in the two competing world systems. Moreover, he coined his own word for supporters of the "Disneyoid picture" of the exploding early universe – "big bangers." According to Russell Hanson, the difference between the big bangers and the "continual creators" was basically of a semantic nature, rooted in different meanings given to words such as creation and universe. To his mind, none of the two world systems was satisfactory from a philosophical point of view. However, he seriously misrepresented the steady state theory by stating that it shared with the big bang theory the view that "our universe, in its very early youth, was considerably different in constitution and appearance from what it is now" (p. 467).

## 6. Microwaves and the big bang

The mid-1960s was a watershed in the history of modern cosmology. While big bang theories of the universe had been proposed earlier – first by Lemaître and later by Gamow and his collaborators – it was only in 1965 that the theory received such spectacular confirmation that it changed its status from outsider to mainstream theory (Brush 1993; Kragh 1996, pp. 318-380). With improved data for the distribution of radio sources and the new quasars, and particularly with the discovery of the cosmic microwave background, steady state cosmology was largely abandoned and left the cosmological scene to the now victorious hot big bang theory. By the end of the decade, this theory, consisting of a large class of models sharing the assumption of a hot and dense beginning of the universe, had become a standard theory accepted by a large majority of physicists and astronomers.

The decisive turn in the status of the big bang occurred with the recognition in the spring of 1965 that an unexplained microwave radiation found by Arno Penzias and Robert Wilson the year before could be understood as a fossil of the exploding universe. Although such a cosmic background radiation had been predicted by Alpher and Herman in 1948, it was independent work by Robert Dicke and James Peebles at Princeton University that led to the conclusion that the Penzias-Wilson radiation was a relic from the primordial decoupling of matter and radiation caused by the early cooling of the expanding universe (for details, see Peebles, Page and Partridge 2009).

Peebles and his Princeton associates did not use the term big bang in their seminal paper in the *Astrophysical Journal* of July 1965, but instead referred to the "primordial fireball" (Dicke et al. 1965). With this term they did not refer to the origin of the universe, but to the state at which matter ceased being in thermal equilibrium with radiation and the universe thus became transparent to radiation. In other words, they referred to the origin of the cosmic background radiation. Only the following year, when calculating the formation of helium in the early universe, did Peebles use the name "big bang" and on this occasion he also referred for the first time to the earlier work of Gamow, Alpher and Herman. He only used the name once, and in citation marks, apparently identifying it with Gamow's "theory of the formation of the elements in the early, highly contracted Universe" (Peebles 1966, p. 542).

Big bang had become part of the cosmologists' vocabulary, but in the decade following the discovery of the microwave background it was not widely used. The *Web of Knowledge* lists only 11 scientific papers in the period 1960-1970 with the name in their titles, followed by 23 papers in the period 1971-1975 (Figure 2). According to the SAO/NASA database, which includes also books, book reviews and more popular literature, the numbers for the two periods are 9 and 30, respectively, and *Scopus* lists 3 and 13 papers in the physical sciences with "big bang" in their article titles.

While the discovery of the cosmic microwave background was generally seen as convincing evidence for the big bang, a minority of physicists and astronomers disagreed. Ted Bastian, a physicist at Cambridge University, criticized "the simple realism that is exhibited by some opponents of the steady state theory," accusing them of assuming "that the big bang is adequately modelled by the explosion of a shrapnel shell" (Bastian 1965). Nonetheless, as the British-American astrophysicist George Burbidge noted with dismay in a paper of 1971, "the big bang bandwagon has gained momentum" (Burbidge 1971, p. 40). This was clearly the case, but the name big bang lacked somewhat behind the bandwagon and was not always used in the same sense. Possibly the first research paper referring to big bang in its title, in March 1966 Stephen Hawking and Roger Tayler of Cambridge University examined the synthesis of helium in anisotropic models of the early universe. They took the big bang theory to mean that "the universe had a singular origin with high temperature and density" (Hawking and Taylor 1966). William McCrea, the former advocate of and contributor to the steady state theory, identified "big bang" with a cosmic space-time singularity such as required by the so-called Penrose-Hawking singularity theorem (McCrea 1970). This was a meaning different from the one held by Peebles, Fowler and several other cosmologists, who associated the big bang with the formation of atomic nuclei in the early universe.

According to the Penrose-Hawking theorem there must be at least one singularity in cosmological models based on classical general relativity. In an important paper of 1970 Hawking and Roger Penrose referred to the big bang in the meaning of a space-time singularity, speaking for instance of "an initial (e.g. 'big bang' type) singularity" (Hawking and Penrose 1970, p. 530). McCrea suggested using the name big bang even if there were several singularities and restrict "the big bang" to the case of a single singularity. Neither he nor Hawking and Penrose mentioned Gamow's theory of element formation. The singularity theorem did not necessarily imply a beginning of the universe in a state of infinite density, since it was generally believed that at exceedingly high energies general relativity would have to be replaced with an unknown theory of quantum gravity. Some physicists tended to identify big bang cosmology with the view that "the Universe has evolved from an earlier state in which conditions were so extreme that the presently known laws of physics were inadequate" (Misner 1969, p, 1329).

The discovery of the cosmic background radiation, and with it the revived big bang universe, was widely reported in American newspapers and popular science magazines. As early as 21 May 1965 – before the discovery had been published in *The Astrophysical Journal* – the *New York Times* included on its front page

Walter Sullivan's report on the heavenly signals and their implications for the big bang universe. "The idea of a universe born 'from nothing' raises philosophical as well as scientific problems," Sullivan informed the readers (Sullivan 1965). Reports in other newspapers and journals, including *Scientific American*, *Sky and Telescope* and *Newsweek*, followed over the next few months (see references in Brush 1993). Most of the articles highlighted the novelty of the amazing picture of a big bang universe, although C. P. Gilmore wrote in *Popular Science* about "Lemaître's 'big bang' theory" (Gilmore 1965). Another article, appearing in *Science News Letter* in June 1965, explained: "The 'big bang' theory holds that some 12 or more billion years ago all matter in the universe was in one place and was spewed outward in every direction by a gigantic explosion" (SNL 1965).

Apart from dealing with the new cosmic background radiation, many of the articles also dealt with the recently discovered and at the time still mysterious quasars.[16] For example, this was the theme in an article in *Popular Mechanics* which repeated the explosion analogy to be found in almost all popular expositions: "The big bang theory postulates that the universe began with a huge explosion of an original mass" (Pearson 1965, p. 234). At about the same time the cosmological big bang picture appeared – probably for the first time – in a commercial context, an advertisement from the Eastman Kodak Company. "Quasars are receding at nearly the speed of light," the advertisement said. "They must be out by the 'shock wave' from the original big bang of creation" (*Science* 149, 17 Sep 1965, p. 1320).

As illustrated by a paper in *Sky and Telescope* written by the astronomer Thornton Page, the connection between quasars and big bang cosmology appeared in popular journals even before the announcement of the discovery of the cosmic microwave background. In early 1965 Page referred several times to the big bang – "the explosion that started the expansion of the universe" – which he related to Lemaître's primeval atom and Gamow's theory of element formation. Moreover, he pointed out that the distribution and distances of the quasars "may tell us whether the Big-Bang or the Steady-State cosmology is correct" (Page 1965, reprinted in Page and Page 1969, pp. 133-150). Indeed, by 1966 analyses of quasar data provided additional convincing evidence that the steady state theory could not be correct.

---

[16]   The name "quasar" was proposed by Hong-Yee Chiu of Columbia University in 1964, but until about 1966 the puzzling objects were generally known as QSOs or quasi-stellar objects. Like it was the case with "big bang," in the beginning "quasars" almost always appeared in inverted commas. On the cosmological significance of quasars and their role in the final phase of the steady state theory, see Kragh (1996, pp. 331-338).

Hoyle tried to evade the conclusion by a drastic modification of the steady state theory, incorporating features of the big bang picture in it: "Perhaps the quasars are an indication that the universe has lots of little bangs instead of one big bang, little bangs that are nevertheless far more violent than the gentle processes of the steady-state theory" (Hoyle 1965, p. 56). The phrase "little bangs" introduced by Hoyle came to be widely used, both in astrophysics and in high energy physics experiments simulating the conditions of the big bang. (Collisions between heavy ions at high energy are sometimes described by the oxymoron "little big bang.") In Hoyle's revision of the steady state theory made in collaboration with Jayant Narlikar and Geoffrey Burbidge, eventually leading to the so-called quasi steady state cosmology, little bangs played an important role.

By the early 1970s the hot big bang theory had acquired a nearly paradigmatic status, and at the same time cosmology experienced a strong quantitative growth. Whereas the annual number of scientific articles on cosmology had on average been about 30 in the period 1950-1962, between 1962 and 1972 the number increased from 50 to 250.[17] An opinion survey of predominantly American

---

[17] For bibliometric illustrations of the growth of cosmology, see Kaiser (2006, p. 447), and Marx and Bornmann (2010, p. 543). However, the growth is in some respect illusory, as the number of publications in the physical and astronomical sciences as a whole grew even more

astronomers conducted in 1959 showed 33% to agree that "the universe started with a 'big bang' several billion years ago." In a later poll of 1980 the figure had increased to 69%, and had it been restricted to astronomers active in cosmological research it would undoubtedly have shown an even greater approval of big bang cosmology. In spite of the growth and the new consensus, cosmology remained a small and loosely organized science.

Another indication of cosmology's growing maturity as a scientific discipline is the emergence of textbooks on the now standard theory of the universe. Peebles' *Physical Cosmology* from 1971, the first textbook with this title, was based on a graduate course he gave in Princeton two years earlier. The term "big bang" appeared frequently in the book, although not as frequently as "primeval fireball," a term he used to designate the cosmic background radiation, either now or at the time of matter-radiation decoupling. Thus, Peebles spoke of the present microwave background at temperature 2.7 K as the primeval fireball – a very cold fireball, then – and he described the consensus model of the universe as the "Big Bang Fireball picture" (Peebles 1971, p. 194 and p. 240). Dennis Sciama's more elementary *Modern Cosmology* from the same year identified the concept of the big bang with the concept originally introduced by Gamow and his group: "The idea that the early dense stages of the Universe were hot enough to enable thermonuclear reactions to occur … has come to be known as the $\alpha$-$\beta$-$\gamma$ theory … or as the hot big bang theory" (Sciama 1971, p. 156).

While the two mentioned textbooks made frequent use of the term big bang, Steven Weinberg – who had used the term as early as 1962 – largely avoided it in his advanced text *Gravitation and Cosmology* from 1972. Speaking instead of the "standard model," a phrase he is likely to have borrowed from his own field of particle physics,

---

rapidly. While cosmology in 1950 made up 0.4% of the physics research papers, in 1970 the percentage had shrunk to a little less than 0.3% (Ryan and Shepley 1976).

only once did he refer to the "big-bang Friedmann model" (Weinberg 1972, p. 611).[18] Yet another important book from the early period, Yakov Zel'dovich and Igor Novikov's encyclopedic *Relativistic Astrophysics*, avoided the term altogether. The two Russian authors based their exposition on what they called the Friedmann theory of a singular beginning of the universe, referring throughout to the "theory of the hot Universe" as an alternative to the hot big bang theory. With the hot universe they meant the original state of "a Universe filled with material that is dominated by photons" (Zel'dovich and Novikov 1983, p. xxv; translation of Russian original of 1975). Neither in their earlier works did they speak of a big bang. Finally, in their massive textbook *Gravitation* of 1973, Charles Misner, Kip Thorne and Wheeler included a book-length chapter on modern cosmology. The three authors referred a few times to the "standard hot big-bang model," using the term big bang as a name for a model rather than an event at the beginning of the cosmic expansion (Misner, Thorne and Wheeler 1973, p. 763).

## 7. Proliferations

In the last quarter of the twentieth century cosmology was completely dominated by versions of the hot big bang, with or without an early inflation era. Hoyle's old name was now commonly used, although the number of scientific papers referring to big bang as part of their topic remained low until about 1990, after which it increased drastically (figures 2 and 3).[19] Until the end of 2012 the *Web of Knowledge* lists a total of 4,548 science papers with "big bang" as their topic, with only 198 of them published before 1990. *Scopus* includes 4,077 papers from 1960-2012 with the name in title, abstract or key words, of which 3,673 are in the physical sciences. The corresponding figures for 1960-1989 are 422 and 404, respectively. Another way of illustrating the popularity of the big bang term is to search for the term in the databases of journals such as *Nature* and *Science*. The result is shown in Figure 4. To evaluate the bibliometric data one has to take into account the general growth in cosmology publications. Between 1980 and 2000 the number of articles on cosmology grew steadily from about 400 per year to a little more than 2,000 per year (Marx and Bornmann 2010, p. 447). The sudden increase of references to "big bang" in the 1990s does not correspond to a similar increase in cosmology publications. Finally, Figure 5 gives data from the arxiv e-print data base, section on astrophysics, from its start in April 1992 to the end of 2012.

---

[18]  In his later and much extended textbook on cosmology, Peebles adopted Weinberg's "standard model" as a less misleading term than "big bang model." As he pointed out, and as several other cosmologists have called attention to, big bang is a misnomer. This is not only because it alludes to a noisy explosion in space but also because the big bang, if taken to be a creation event or the event when the universe started expanding, is outside the standard models of cosmology and physics (Peebles 1993, p. 6).

[19]  It is not clear what the *Web of Knowledge* means by "topic," and there are other reasons why information from this database should be used with some caution.

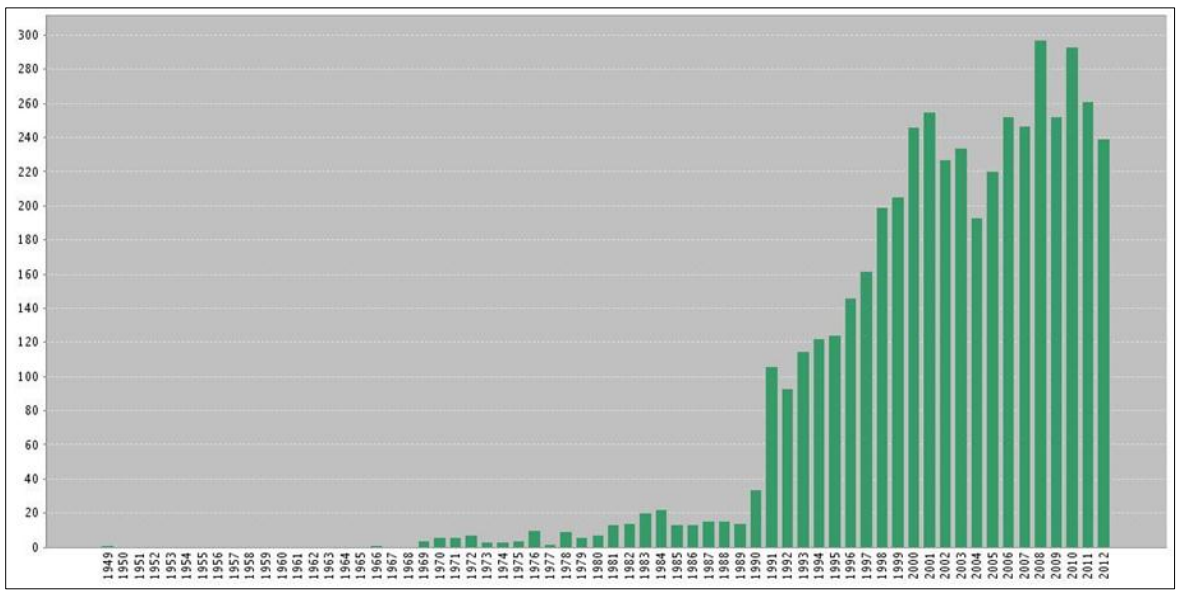

Figure 2. *Web of Knowledge* data showing the number of science papers with "big bang" as their topic (assessed 31 December 2012). Total number = 4,548.

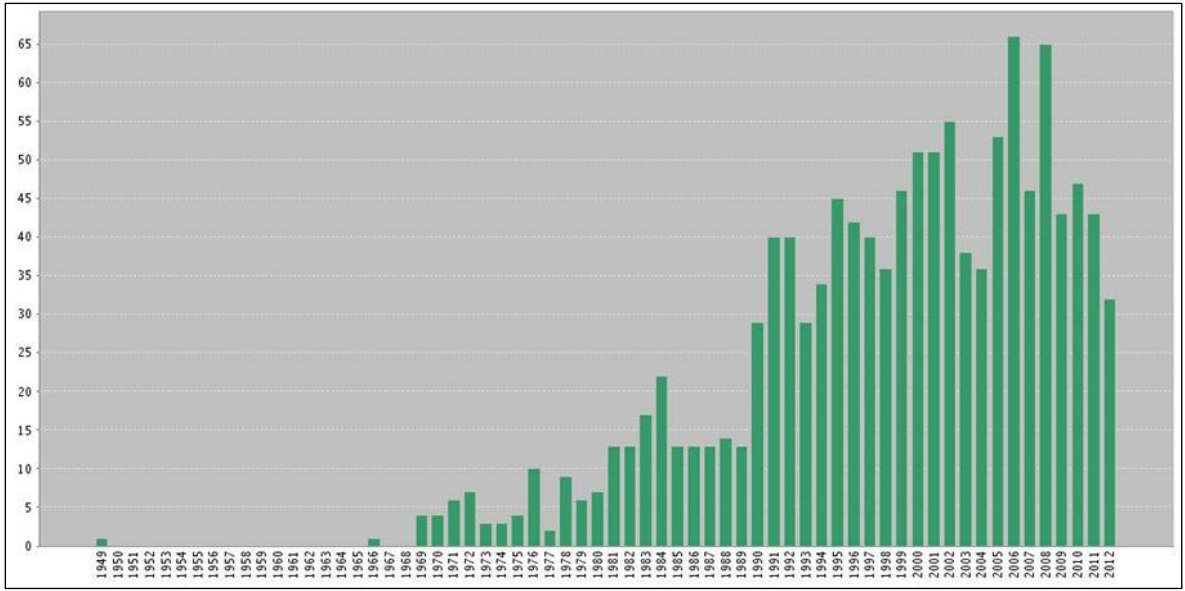

Figure 3. *Web of Knowledge* data showing the number of science papers with "big bang" appearing in their title (assessed 31 December 2012). Total number = 1,205. The corresponding number in the SAO/NASA database (astronomy and physics) is 1,837 or, including arxiv e-prints, 1,960.

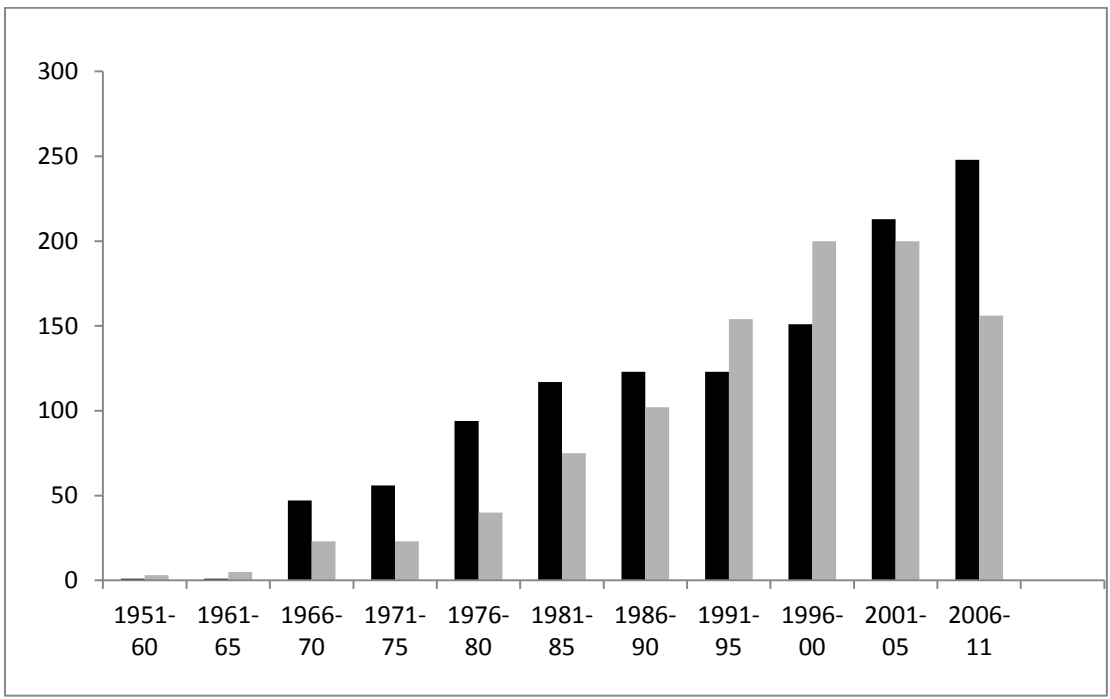

Figure 4. Number of articles or notes 1951-2011 in *Nature* (black) and *Science* (grey) with references to big bang. Not all the references are to cosmology, but the large majority is.

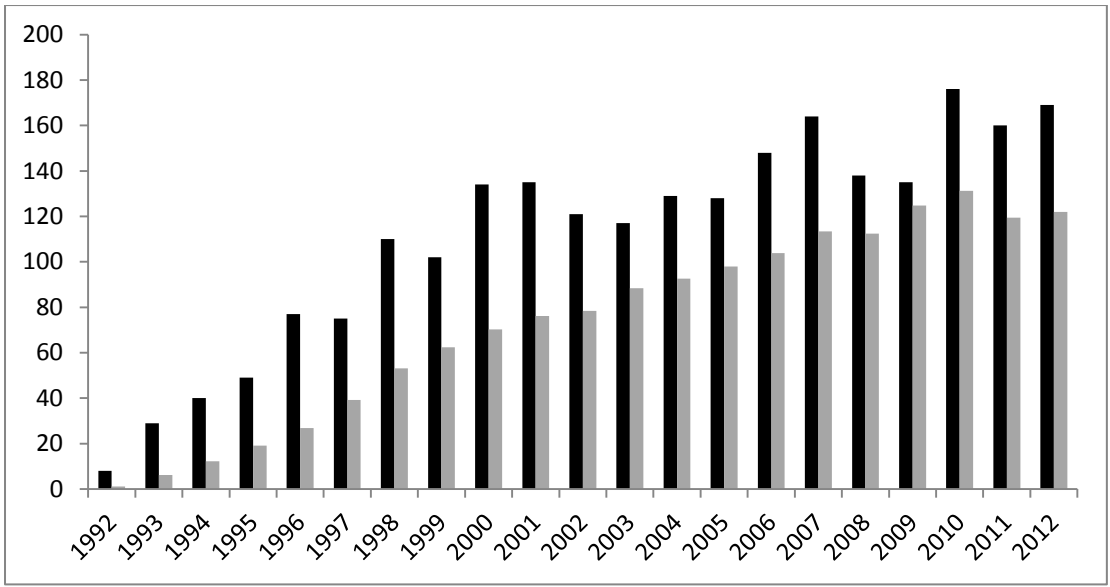

Figure 5. Number of papers in the arxiv astrophysics database 1992-2012 with "big bang" in their abstracts (black). The grey columns are the total number of papers divided by 100. (http://arxiv.org/archive/astro-ph).

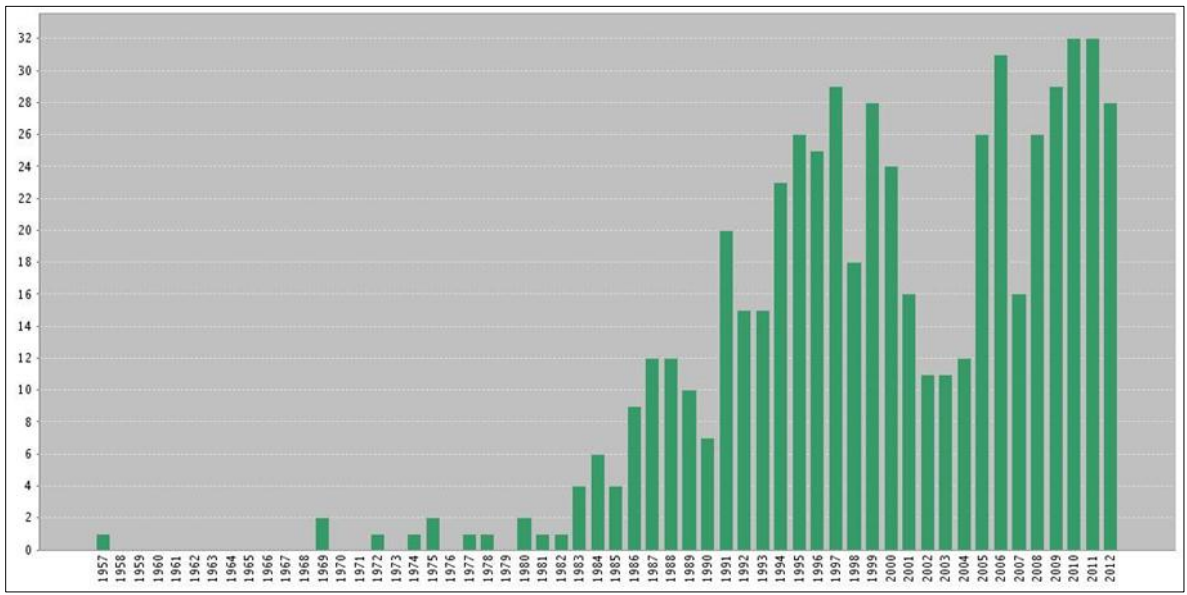

Figure 6. *Web of Knowledge* data showing the number of papers in humanities, arts and social sciences with "big bang" as their topic (assessed 31 December 2012). Total number = 498.

Probably as a result of the popularity of the name in cosmology, and of cosmology's wide public appeal, "big bang" began to appear in many other contexts, both within and without the sciences. Thus, about 500 papers referring to big bang have been published in the humanities and the social sciences (Figure 6). The number of articles in this category with the name in their titles is about 200.

In some cases "big bang" is used in a general sense to characterize a sudden, comprehensive and drastic change rather than one which occurs incrementally and over long spans of time. The Tunguska event in 1908 has been described as Siberia's big bang (Rich 1978), and biologists sometimes speak of the sudden appearance of life forms in the Cambrian era almost 600 million years ago as biology's big bang (Levinton 1992). Such uses of the term in the biological literature may allude explicitly to cosmology, as in a paper entitled "Origin of Genes: 'Big Bang' or Continuous Creation?" (Keese and Gibbs 1992; see also Koonin 2007). Similar metaphorical kinds of usage are quite common today also in other areas, not least in economic and financial studies, where it goes back to the 1980s. For example, the big bang metaphor has been used extensively in discussions of how to transform centrally planned economies into market-oriented ones, as in the cases of China and Eastern Europe. As one author explains, "A big-bang or shock therapy approach implements various reforms … quickly, in a concentrated time frame, whereas a gradualist approach spreads various reforms over an extended period" (Wei 1997, p. 1234). The jumps of Poland and Hungary to market economies in the early 1990s were described as economic big bangs, with some economists warning that the big bang approach might be damaging (Hare and Révész 1992).

Finally, the label "big bang" is today used in a variety of commercial, cultural and artistic contexts that has only the name in common with the original cosmological meaning of the term. Numerous music albums, television series, films, comics, sport events and commercial products of any kind carry the name that Hoyle casually coined in 1949. Rolling Stone's album of 2005 called "A Bigger Bang" is one example, the popular American television sitcom "Big Bang Theory" is another. A Google search for "big bang" gives 167 million results, while the search on Google Scholar gives 231,000 results (the combination "big bang" and "universe" reduces the hits to 28.6 million and 124,000, respectively). Although it took a long time for the big bang to become catchy, it definitely has become so.

## 8. Conclusions

Names are important, in science as elsewhere. An apt name for an object, phenomenon or concept may help to bring it recognition and visibility, or to associate it with certain mental images that other names do not convey. "Big bang" is one such name that today is almost synonymous with modern cosmology, but the success of the name is of relatively modern date. A detailed account of the origin and development of the label referring to the beginning of the universe adds some new information to the history of cosmological thought in the period from about 1930 to 1970 out of which the modern view of the universe arose. Contrary to what is usually assumed, one can find the name big bang in scientific contexts (meteorology) even before Hoyle introduced it in cosmology in 1949. Strictly speaking Hoyle did not invent the term, he reinvented it in a new context and gave it a new meaning.

Far from being an instant success, for a long time the later so famous name was ignored by most physicists and astronomers, including Hoyle himself. It was used earlier and more commonly in popular than in scientific publications, and more in the United States than in England. Only in the 1970s began "big bang" to be commonly used in cosmology, and it took more than another decade until it really caught on and came into wide use also outside the small field of physical cosmology. One might imagine that the "big bang" label helped making the idea of a finite-age universe popular and in this way contributed to the victory of this class of models in the mid-1960s. But this is not what happened. The wide acceptance of the label occurred post hoc, only after observations had provided convincing evidence that we live in a big bang universe. In other words, there is no reason to assume that Hoyle's name had a causal effect on the development of cosmology in the period.

This paper has called attention to several misunderstandings concerning the origin and early history of the term big bang and its associated concept. One of them is that the term was due to Gamow, or that Gamow promoted it in order to support his favoured conception of the early universe. In reality, Gamow disliked the term and almost never used it. Another popular misconception is that Hoyle coined the term to ridicule or disparage theories with a definite origin of the universe, such as Gamow's. Although one cannot rule it out, there is no documentation in the historical records for this often repeated view. Although one cannot rule it out, this often repeated view lacks convincing evidence. Had Hoyle intended the term in this way, he would presumably have used it frequently during the cosmological controversy, which he did not. Between 1950 and 1965 he did not refer to Big Bang at all.

Names usually have several connotations and meanings, such as it was the case with the big bang in the early period and such as is still the case. It is not a well-defined concept and has never been, except that it refers to some kind of original cosmic state out of which our present universe arose. When physicists and astronomers spoke of the big bang in the 1960s, it was basically in two different ways. According to one view, the big bang was a cosmic space-time singularity in the strict sense, an absolute beginning corresponding to an unphysical point of infinite density (Heckmann 1961; Hawking and Penrose 1970; McCrea 1970). On the other hand, it was more common to associate the initial event with a physical state of the kind imagined by Gamow or Lemaître, a primordial high-density and high-temperature universe made up of radiation and nuclear particles in a state of rapid

expansion – often described as an explosion (Fowler 1957; Page 1965; Sciama 1971). A few physicists identified the big bang with a state so early and exotic that it could not be understood in terms of the known laws of physics (Misner 1969).

These are images still found today, both among cosmologists and in the general public. In part because the big bang invites images of a primordial event in the form of an explosion, it is widely agreed that the term is a misnomer and that it causes more confusion than clarity. It certainly does in the teaching of science, where studies consistently show that students tend to associate the big bang with an explosion of pre-existing matter into empty space (Prather, Slater and Offerdahl 2003). And no wonder, for the explosion metaphor is still routinely used in the popular literature, much as it was in the past. Perhaps big bang is a misnomer, but as Hoyle commented, it is one of those catchy harpoon-like words that are very hard to pull out once it has gone in.

## References

AIP (1968). Interview with George Gamow by Charles Weiner, 25 April 1968, Niels Bohr Library & Archives, American Institute of Physics (http://www.aip.org/history/ohilist/4325.html)

Alfvén, Hannes (1966). *Worlds-Antiworlds: Antimatter in Cosmology*. San Francisco: W. H. Freeman.

Alpher, Ralph and Robert Herman (1969). "Reflections on 'big bang' cosmology," *General Electric Research and Development Center*, Report No. 69-C-165.

Alpher, Ralph and Robert Herman (1988). "Reflections on early work on 'big bang' cosmology," *Physics Today* 41: 8, 24-34.

Alpher, Ralph and Robert Herman (1990). "Early work on 'big-bang' cosmology and the cosmic background radiation," pp. 129-158 in Bruno Bertotti et al., eds., *Modern Cosmology in Retrospect*. Cambridge: Cambridge University Press.

Alpher, Ralph and Robert Herman (1997). "Cosmochemistry and the early universe," pp. 49-70 in Eamon Harper, W. C. Parke and G. D. Anderson, eds., *The George Gamow Symposium*. San Francisco: Astronomical Society of the Pacific.

Baldwin, Hanson W. (1956). "The new face of war," *Bulletin of Atomic Scientists* 12: 5, 153-158.

Barnes, Ernest W. (1933). *Scientific Theory and Religion: The World Described by Science and its Spiritual Interpretation*. Cambridge: Cambridge University Press.

Bastian, Ted (1965). "For and against the steady state," *New Scientist* 27, 164.

Beatty , Cheryl J. and Richard T. Fienberg (1994). "Participatory cosmology: The big bang challenge," *Sky and Telescope* 87: 3, 20-22.

Belenkiy, Ari (2012). "Alexander Friedmann and the origins of modern cosmology," *Physics Today* 65: 10, 38-43.

Bonnor, William B. (1964). *The Mystery of the Expanding Universe*. New York: Macmillan.

Brush, Stephen G. (1993). "Prediction and theory evaluation: Cosmic microwaves and the revival of the big bang," *Perspectives on Science* 1, 245-278.

Burbidge, Geoffrey R. (1971). "Was there really a big bang?" *Nature* 233, 36-40.

Chernin, Arthur D. (1995). "George Gamow and the big bang," *Space Science Reviews* 74, 447-454.

Cox, E. F. et al. (1949). "Upper-atmosphere temperatures from Helgoland big bang," *Journal of Meteorology* 6, 300-311.

Deunk, Neil H. (1962). "A controversy in contemporary cosmology," *School Science and Mathematics* 62: 7, 487-502.

Dicke, Robert H., P. James E. Peebles, Peter G. Roll and David T. Wilkinson (1965). "Cosmic black-body radiation," *Astrophysical Journal* 142, 414-419.

Dyson, Freeman J. (1959). "The future development of nuclear weapons," *Foreign Affairs* 38, 457-464.

Economist 1957. "The big bang," *The Economist*, 2 February 1957, 365.

Eddington, Arthur (1928). *The Nature of the Physical World*. Cambridge: Cambridge University Press.

Eddington, Arthur (1931). "The expansion of the universe," *Monthly Notices of the Royal Astronomical Society* 91, 412-416.

Einstein, Albert (1929). "Space-time," pp. 105-108 in *Encyclopedia Britannica* (London: Encyclopedia Britannica, Inc.), 14th ed., vol. 21.

Ferris, Timothy (1977). *The Red Limit: The Search for the Edge of the Universe*. New York: William Morrow and Co.

Ferris, Timothy (1993). "Needed: A better name for the big bang," *Sky and Telescope* 86: 2, 4-5.

Fölsing, Albert (1997). *Albert Einstein*. New York: Penguin Books.

Fowler, William A. (1956). "The origin of the elements," pp. 17-31 in Gerard Piel et al., eds., *The Universe*. New York: Simon and Schuster.

Fowler, William A. (1957). "Formation of the elements," *Scientific Monthly* 84, 84-100.

Friedmann, Alexander (1922). "Über die Krümmung des Raumes," *Zeitschrift für Physik* 10, 377-386.

Gamow, George (1940). *The Birth and Death of the Sun*. New York: Viking Press.

Gamow, George (1952). *The Creation of the Universe*. New York: Viking Press.

Gamow, George (1956). "The evolutionary universe," *Scientific American* 192: 9, 136-154.

Gamow, George (1961). "Gravity," *Scientific American* 204: 3, 94-106.

Gamow, George (1970). *My World Line: An Informal Autobiography*. New York: Viking Press.

Gilmore, C. P. (1965). "They're solving the world's greatest mystery," *Popular Science* 187 (November), 102-105, 200-203.

Gregory, Jane (2005). *Fred Hoyle's Universe*. Oxford: Oxford University Press.

Hanson, Norwood Russell (1963). "Some philosophical aspects of contemporary cosmologies," pp. 227-264 in Bernard H. Baumrin, ed., *Philosophy of Science: The Delaware Seminar*, vol. 2. New York: Interscience.

Hare, Paul and Tamas Révész (1992). "Hungary's transition to the market: The case against a 'big bang'," *Economic Policy* 7, 227-264.

Harper, Eamon (2001). "George Gamow: Scientific amateur and polymath," *Physics in Perspective* 3, 335-372.

Harrison, Edward R. (1968). "Lagging cores and little bangs," *Astronomical Journal* 73, S182.

Hawking, Stephen W. and Roger Penrose (1970). "The singularities of gravitational collapse and cosmology," *Proceedings of the Royal Society of London A* 314, 529-548.

Hawking, Stephen W. and Roger J. Tayler (1966). "Helium production in an anisotropic big-bang cosmology," *Nature* 209, 1278-1279.

Heckmann, Otto (1961). "On the possible influence of a general rotation on the expansion of the universe," *Astronomical Journal* 66, 599-603.

Horgan, John (1995). "The return of the maverick," *Scientific American* 272 (March), 46-47.

Hospers, John (1953). *An Introduction to Philosophical Analysis*. New York: Prentice Hall.

Hoyle, Fred (1949). "Continuous creation," *The Listener* 41, 567-568.

Hoyle, Fred (1950a). *The Nature of the Universe*. Oxford: Blackwell.

Hoyle, Fred (1950b). *The Nature of the Universe*. New York: Harper & Brothers.

Hoyle, Fred (1956). "The steady-state universe," *Scientific American* 192: 9, 157-166.

Hoyle, Fred (1961). "Observational tests in cosmology," *Proceedings of the Physical Society* 77, 1-16.

Hoyle, Fred (1965). *Galaxies, Nuclei, and Quasars*. New York: Harper & Row.

Israel, Werner (1987). "Dark stars: The evolution of an idea," pp. 199-276 in S. Hawking and W. Israel, eds., *Three Hundred Years of Gravitation*. Cambridge: Cambridge University Press.

Jordan, Pascual (1936). *Die Physik des 20. Jahrhunderts*. Braunschweig: Vieweg.

Jordan, Pascual (1937). "Die physikalischen Weltkonstanten," *Die Naturwissenschaften* 25, 513-517.

Kaiser, David (2006). "Whose mass is it anyway? Particle cosmology and the objects of theory," *Social Studies of Science* 36, 533-564.

Kaiser, David (2007). "The other evolution wars," *American Scientist* 95, 518-525.

Keese, Paul K. and Adrian Gibbs (1992). "Origins of genes: 'Big bang' or continuous creation?" *Proceedings of the National Academy of Science* 89, 9489-9493.

Koonin, Eugene V. (2007). "The biological big bang model for the major transitions in evolution," *Biology Direct* 2, 1-17.

Kragh, Helge (1996). *Cosmology and Controversy. The Historical Development of Two Theories of the Universe*. Princeton: Princeton University Press.

Kragh, Helge and Dominique Lambert (2007). "The context of discovery: Lemaître and the origin of the primeval-atom universe," *Annals of Science* 64, 445-470.

Kragh, Helge and Stephen J. Weininger (1996). "Sooner silence than confusion: The tortuous entry of entropy into chemistry," *Historical Studies in the Physical and Biological Sciences* 27, 91-130.

Lemaître, Georges (1931). "The beginning of the world from the point of view of quantum theory," *Nature* 127, 706.

Lemaître, Georges (1946). *L'Hypothèse de l'Atome Primitif: Essai de Cosmogonie*. Neuchâtel: Éditions du Griffon.

Lemaître, Georges (1950). *The Primeval Atom: A Hypothesis of the Origin of the Universe*. New York: Van Nostrand.

Lemaître, Georges (1958). "The primeval atom hypothesis and the problem of the clusters of galaxies," pp. 1-32 in R. Stoops, ed., *La Structure et l'Évolution de l'Univers*. Brussels: Coudenberg.

Levinton, Jeffrey (1992). "The big bang of animal evolution," *Scientific American* 267 (November), 84-91.

Lightman, Alan and Roberta Brawer (1990). *Origins: The Lives and Worlds of Modern Cosmologists*. Cambridge, Mass.: Harvard University Press.

Lyttleton, Raymond A. (1956). *The Modern Universe* (London: Hodder & Stoughton).

Lyttleton, Raymond A. (1959). "Man and the universe," *Nature* 183, 1624-1625.

Mann, Martin (1962). "The march of science: Big-bang up, steady-state down," *Popular Science* 180: 3, 29.

Marx, Werner and Lutz Bornmann (2010). "How accurately does Thomas Kuhn's model of paradigm change describe the transition from the static view of the universe to the big bang theory in cosmology?" *Scientometrics* 84, 441-464.

Mather, Kirtley F. (1951). "Current science reading," *Science* 113, 427-429.

McCafferty, Phil (1963). "Big bang for little bullets," *Popular Science* 182: 1, 112-115.

McConnell, Craig S. (2006). "The BBC, the Victoria Institute, and the theological context for the big bang-steady state debate," *Science & Christian Belief* 18, 151-168.

McCrea, William H. (1963). "Cosmological theories – a survey," pp. 206-221 in H. P. Palmer, R. D. Davies and M. I. Large, eds., *Radio Astronomy Today*. Manchester: Manchester University Press.

McCrea, William H. (1970). "A philosophy for big-bang cosmology," *Nature* 228, 21-24.

McVittie, George C. (1951). "The cosmological problem," *Science News* 21, 61-75.

McVittie, George C. (1961). "Rationalism versus empiricism in cosmology," *Science* 133, 1231-1236.

McVittie, George C. (1974). "Distance and large redshifts," *Quarterly Journal of the Royal Astronomical Society* 15, 246-263.

Menzel, Donald H. (1932). "Blast of giant atom created our universe," *Popular Science* 105, 28-29.

Misner, Charles W. (1969). "Absolute zero of time," *Physical Review* 186, 1328-1333.

Misner, Charles W., Kip S. Thorne and John A. Wheeler (1973). *Gravitation*. San Francisco: W. H. Freeman.

Mitton, Simon (2005). *Fred Hoyle. A Life in Science*. London: Aurum.

Nussbaumer, Harry and Lydia Bieri (2009). *Discovering the Expanding Universe*. Cambridge: Cambridge University Press.

O'Connell, Daniel (1952). "According to Hoyle," *Irish Astronomical Journal* 2, 627-648.

Page, Thornton (1965). "The evolution of galaxies," *Sky and Telescope* 29 (January-February), 4-7, 81-84.

Page, Thornton and Lou W. Page, eds. (1969). *Beyond the Milky Way: Galaxies, Quasars, and the New Cosmology*. New York: Macmillan.

Pearson, John F. (1965). "At the edge of space," *Popular Mechanics* 124: 3, 94-97, 232-238.

Peebles, P. James E. (1966). "Primordial helium abundance and the primordial fireball, II," *Astrophysical Journal* 146, 542-552.

Peebles, P. James E. (1971). *Physical Cosmology*. Princeton: Princeton University Press.

Peebles, P. James E. (1984). "Impact of Lemaître's ideas on modern cosmology," pp. 23-30 in A. Berger, ed., *The Big Bang and Georges Lemaître*. Dordrecht: Reidel.

Peebles, P. James E. (1993). *Principles of Physical Cosmology*. Princeton: Princeton University Press.

Peebles, P. James E., Lyman A. Page and R. Bruce Partridge (2009). *Finding the Big Bang*. Cambridge: Cambridge University Press.

Pfeiffer, John (1956). *The Changing Universe: The Story of the New Astronomy*. London: Victor Gollancz.

Prather, Edward E., Timothy F. Slater and Erika G. Offerdahl (2003). "Hints of a fundamental misconception in cosmology," *Astronomy Education Review* 1: 2.

Raychaudhury, Somak (2004). "And Gamow said, let there be a hot universe," *Resonance* 10 (July), 220-231.

Rich, Vera (1978). "The 70-year-old mystery of Siberia's big bang," *Nature* 274, 207.

Ryan, Michael P. and Shepley, L. C. (1976). "Resource letter RC-1: Cosmology," *American Journal of Physics* 44, 223-230.

Ryle, P. J. (1924). "Study of explosions," *Nature* 114, 123.

Sanchez-Ron, José M. (2005). "George McVittie, the uncompromising empiricist," pp. 189-222 in Anne J. Kox and Jean Eisenstadt, eds., *The Universe of General Relativity* (Boston: Birkhäuser).

Schneider, Barry (1975). "Big bangs from little bombs," *Bulletin of Atomic Scientists* 31: 5, 24-29.

Schröder, Wilfried and Hans-Jürgen Treder (1996). "Hans Ertel and cosmology," *Foundations of Physics* 26, 1081-1088.

Sciama, Dennis (1971). *Modern Cosmology*. Cambridge: Cambridge University Press.

Sidhart, B. G. and Rhawn Joseph (2010). "Different routes to multiverses and an infinite universe," *Journal of Cosmology* 4, 641-654 (http://journalofcosmology.com/Multiverse8.html).

Singh, Simon (2004). *Big Bang*. London: Fourth Estate.

Smirnov, Yuri N. (1965). "Hydrogen and He-4 formation in the prestellar Gamow universe," *Soviet Astronomy-AJ* 8, 864-867.

Smoot, George and Keay Davidson (1994). *Wrinkles in Time*. New York: William Morris & Company.

SNL (1948). "Early stages of the universe," *Science News Letter* 53: 1, 259.

SNL (1965). "'Big bang' theory upheld," *Science News Letter* 88 (26 June), 403.

Struve, Otto and Velta Zebergs (1962). *Astronomy of the 20th Century*. New York: Macmillan.

Stuewer, Roger H. (1986). "The naming of the deuteron," *American Journal of Physics* 54, 206-218.

Sullivan, Walter (1965). "Signals imply a 'big bang' universe," *New York Times* (21 May).

Thorne, Kip S. (1994). *Black Holes and Time Warps: Einstein's Outrageous Legacy*. New York: Norton.

Trimble, Virginia (2000). "The first explosions," pp. 3-21 in Stephen S. Holt and William W. Zhang, eds., *Cosmic Explosions: Tenth Astrophysical Conference*. College Park, MD: American Institute of Physics.

Wei, Shang-Jin (1997). "Gradualism versus big bang: Speed and sustainability of reforms," *Canadian Journal of Economics* 30, 1234-1247.

Weinberg, Steven (1962). "Universal neutrino degeneracy," *Physical Review* 128, 1457-1473.

Weinberg, Steven (1972). *Gravitation and Cosmology: Principles and Applications of the General Theory of Relativity*. New York: Wiley.

Wheeler, John A. (1968). "Our universe: The known and the unknown," *American Scientist* 56: 1, 1-20.

Whipple, Francis J. W. (1926). "Audibility of explosions and the constitution of the upper atmosphere," *Nature* 118, 309-313.

Williamson, Ralph E. (1951). "Fred Hoyle's universe," *Journal of the Royal Astronomical Society of Canada* 45, 185-189.

Zel'dovich, Yakov B. and Igor D. Novikov (1983). *Relativistic Astrophysics, II: The Structure and Evolution of the Universe*. Chicago: University of Chicago Press.